\documentclass{article}

\usepackage{arxiv}
\usepackage{amsmath,amssymb,amsfonts}
\usepackage[linesnumbered,ruled,vlined]{algorithm2e}
\usepackage{graphicx}
\usepackage{textcomp}
\usepackage{enumerate}
\usepackage{xcolor}
\usepackage{ntheorem}
\usepackage{cite}
\usepackage{stfloats}
\usepackage{array}
\usepackage{bbding}
\usepackage{url}
\usepackage[flushleft]{threeparttable}
\usepackage{verbatim}
\usepackage{booktabs} 
\usepackage{makecell}
\usepackage{listings}
\usepackage{float}
\usepackage{subfig}

\usepackage[utf8]{inputenc} 
\usepackage[T1]{fontenc}    
\usepackage{hyperref}       
\usepackage{nicefrac}       
\usepackage{microtype}      
\usepackage{lipsum}

\title{{AirFogSim: A Light-Weight and Modular Simulator for UAV-Integrated Vehicular Fog Computing}}
\author{
    Zhiwei Wei \\
    Shanghai Research Institute for Intelligent Autonomous Systems\\
    Tongji University\\
    Shanghai, 201210, China\\
    \texttt{2311769@tongji.edu.cn} \\
    \And
    Chenran Huang \\
    School of Software Engineering\\
    Tongji University\\
    Shanghai, 200092, China\\
    \texttt{hcr@tongji.edu.cn} \\
    \And
    Bing Li\\
    School of Software Engineering\\
    Tongji University\\
    Shanghai, 200092, China\\
    \texttt{lizi@tongji.edu.cn} \\
    \And
    Yiting Zhao \\
    School of Software Engineering\\
    Tongji University\\
    Shanghai, 200092, China\\
    \texttt{2153401@tongji.edu.cn} \\
    \And
    Xiang Cheng\\
    School of Electronics\\
    Peking University\\
    Beijing, 100871, China\\
    \texttt{xiangcheng@pku.edu.cn} \\
    \And
    Liuqing Yang\\
    Internet of Things Thrust \& Intelligent Transportation Thrust\\
    The Hong Kong University of Science and Technology (Guangzhou)\\
    Guangzhou, 511458, China\\
    \texttt{lqyang@ust.hk}\\
    \And
    Rongqing Zhang\\
    School of Software Engineering\\
    Tongji University\\
    Shanghai, 200092, China\\
    \texttt{rongqingz@tongji.edu.cn} 
}

\begin{document}
\maketitle
\begin{abstract}
Vehicular Fog Computing (VFC) is significantly enhancing the efficiency, safety, and computational capabilities of Intelligent Transportation Systems (ITS), and the integration of Unmanned Aerial Vehicles (UAVs) further elevates these advantages by incorporating flexible and auxiliary services. This evolving UAV-integrated VFC paradigm opens new doors while presenting unique complexities within the cooperative computation framework. Foremost among the challenges, modeling the intricate dynamics of aerial-ground interactive computing networks is a significant endeavor, and the absence of a comprehensive and flexible simulation platform may impede the exploration of this field. Inspired by the pressing need for a versatile tool, this paper provides a lightweight and modular aerial-ground collaborative simulation platform, termed \texttt{AirFogSim}. We present the design and implementation of AirFogSim, and demonstrate its versatility with five key missions in the domain of UAV-integrated VFC. A multifaceted use case is carried out to validate AirFogSim's effectiveness, encompassing several integral aspects of the proposed AirFogSim, including UAV trajectory, task offloading, resource allocation, and blockchain. In general, AirFogSim is envisioned to set a new precedent in the UAV-integrated VFC simulation, bridge the gap between theoretical design and practical validation, and pave the way for future intelligent transportation domains. Our code will be available at \url{https://github.com/ZhiweiWei-NAMI/AirFogSim}.
\end{abstract}


\section{Introduction}\label{sec:intro}
The advent of Intelligent Transportation Systems (ITS) represents a monumental shift in the landscape of urban mobility, driven by a growing need for safer and more convenient transportation modes. Central to this transformation is the emergence of Connected and Autonomous Vehicles (CAVs), which epitomize the integration of cutting-edge technology with traditional vehicular networks\cite{wireless_era_X_Cheng}. The main distinguishing features of CAVs are the availability of various onboard sensors (e.g., cameras, LiDAR, radar, etc.) that generate massive amounts of data for perception and decision-making, as well as the enhancement of Vehicle-to-Everything (V2X) communications to interact with other entities. Nonetheless, this powerful combination of computation and communication capabilities is supported by the large amount of (3 to 40 GBit/s per CAV according to Tuxera\cite{blog_cavdata}) data generation, exchange, and processing. As we embrace the concept of the metaverse into vehicular networks \cite{AIGC_vehicular_metaverse_Xu}, more computation-intensive technologies such as VR, AR, and Mixed Reality (MR) are beginning to intersect with autonomous driving. 
Therefore, the next-generation ITS is witnessing a tangible development that demands sophisticated computation, high-bandwidth communication, and seamless collaboration among various network entities. 

In response to these technological demands, Vehicular Fog Computing (VFC) has emerged as a crucial enabler within ITS. By decentralizing data processing and bringing computational resources closer to the edge of the network, VFC significantly reduces the latency associated with cloud-based processing and enhances the overall responsiveness of the system. VFC also proposes a fascinating incentive mechanism, where intelligent vehicles (including both moving and parked vehicles) with idle resources are motivated to serve as vehicular fog nodes\cite{foggy_Muham}. Hereby, given the many-to-many matching dynamics between tasks and resources, the crux of VFC lies in adeptly managing the time-varying and distributed nature of computational tasks, focusing on the crucial but complicated \emph{computation offloading}, which has garnered considerable attention recently\cite{folo_zhu, contract_matching_zhou,trust_comp, ocvc,vfc_priority_Jinming_Shi,traffic_load_vec,redundant_resource,se_vfc,large_scale_vfc,madrl}. 

\begin{figure*}
    \centering
    \includegraphics[width=0.9\textwidth]{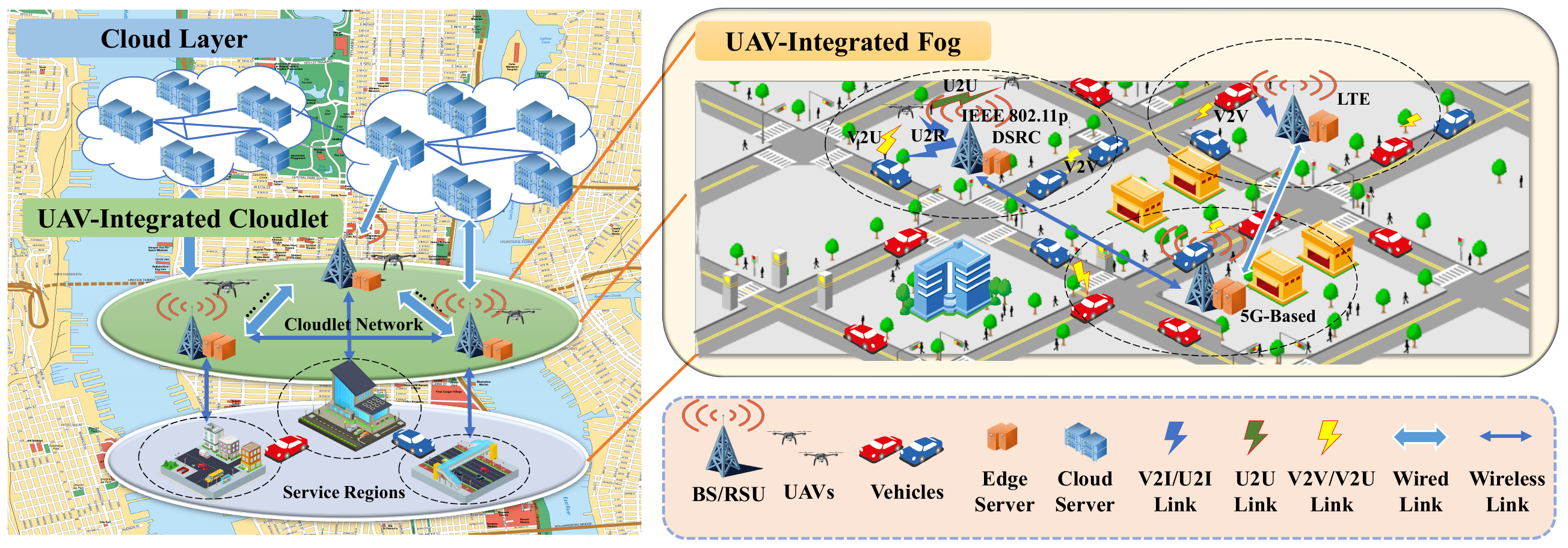}
    \caption{The UAV-integrated vehicular fog computing paradigm.}
    \label{fig:vfc_framework}
    \vspace{-0.4 cm}
\end{figure*}

Though the VFC paradigm greatly alleviates the burden at the static roadside units (RSUs), pervasive communication and computing needs are still outpacing the capabilities of terrestrial vehicular networks. In the coming era, terrestrial vehicular networks and aerial infrastructures are expected to be integrated to provide more ubiquitous wireless connectivity and computing services. For now, divergent modules are mounted on Unmanned Aerial Vehicles (UAVs) to offer more efficient and flexible edge computing, and many researchers \cite{disaster_Y_Wang,coverage_uav,highway_uav,joint_Y_Liu} have begun to explore the potential of UAV-integrated vehicular fog computing (as shown in Fig. \ref{fig:vfc_framework}). On the one hand, UAVs can be deployed as aerial base stations to provide ubiquitous coverage and seamless connectivity, especially in remote areas where terrestrial RSUs are not available. On the other hand, UAVs can be leveraged as mobile fog nodes to offload computation tasks from vehicles, reducing the burden on the ground fog nodes and improving the overall system performance. 
However, the integration of UAVs into vehicular networks is alongside a host of challenges, including the control signal hijacking prevention\cite{blockchain_evastrop}, limited energy, and real-time trajectory planning. Consequently, research into integration mechanisms for computation offloading is essential to fully realize the potential of UAV-integrated VFC.

While research in UAV-integrated VFC is burgeoning, a fundamental challenge persists: how to effectively and accurately model this intricate system and reliably evaluate the algorithm performance? Real-world simulations are often impractical due to the large scale of vehicles and the prohibitive costs associated with traffic disruption, so researchers resort to using simulators with synthetic or real-world data to validate their propositions. Nonetheless, the current landscape of simulation tools \cite{ifogsim,ifogsim2,edgecloudsim,fognetsim++, veins,vfogsim,marsim,skywalker} reveals a critical gap. Many of these tools are either overly specialized, failing to cover the extensive exploration of research interests in the UAV-integrated ITS, or are bogged down by complexity that undermines their practical use in research. There lacks a comprehensive and user-friendly simulation platform, which represents a significant barrier restricting the full exploration for computation offloading in the UAV-integrated VFC. 

This paper aims to address this gap. According to our comprehensive survey on the realm of UAV-integrated VFC, current simulators are either not standardized, limiting their applicability and interoperability, or are excessively cumbersome, posing challenges for efficient deployment and development. In response, our work proposes a light-weight and modular UAV-integrated aerial-ground collaborative vehicular fog computing simulation platform, named as \textbf{AirFogSim}. The constructed platform is unique in its modular design, enabling it to effectively simulate a variety of missions essential to the aerial-ground computation ecosystem. Moreover, the platform's modularity allows for easy extension to additional missions, making it a developing tool for a wide range of research applications in ITS. By providing this comprehensive platform, we aim to empower researchers and engineers to explore new frontiers in the UAV-integrated aerial-ground collaborative VFC, facilitating the development of innovative solutions in intelligent transportation systems. The main contributions of this paper are summarized as follows:

\begin{enumerate}
    \item In this work, we construct the AirFogSim, a versatile, lightweight, and modular simulation platform crafted for computation offloading in UAV-integrated aerial-ground collaborative VFC. This platform is meticulously aligned with contemporary research directions and delivers a comprehensive simulation environment adept at representing the complexities of the aerial-ground interactions. The design of AirFogSim incorporates a selection of current established standards\cite{3gpp_36777,winner2_model,3gpp_36885,3gpp_r15}, thereby enhancing the accuracy and practicability of the platform.
    \item The AirFogSim supports five key missions in the UAV-integrated VFC, which are the RSU/Aerial base station (ABS) deployment, UAV trajectory planning, V2X task offloading, security and privacy, and dynamic resource allocation. By applying various network scheduler modules, the platform simulates the intricate communicating and computing interactions among vehicles, UAVs, and RSUs, and thus enables further research and development by exposing APIs.
    \item To demonstrate the practical utility and effectiveness of AirFogSim, we introduce a detailed use case of the platform. This use case involves the simulation of a UAV-integrated reliable V2X task offloading framework using the blockchain technology. The simulation results verify the platform's validity and demonstrate its capability to accurately model and analyze these complex interactions in UAV-integrated VFC scenarios.
\end{enumerate}

The rest of this paper is concluded as below: Section \ref{sec:related_work} illustrates the background and related work. Section \ref{sec:system_architecture} introduces the system architecture of AirFogSim. Section \ref{sec:example_scenarios} presents five key missions supported by the AirFogSim. Section \ref{sec:simulator} describes the implementation and modeling of different functionalities. Section \ref{sec:use_case} presents a practical use case. Section \ref{sec:conclusion} concludes this paper and proposes future research directions.

\section{Background and Related Work}\label{sec:related_work}
In this section, we first introduce the architecture of UAV-integrated VFC. Then, we summarize the existing research in UAV-integrated VFC and the current simulators.

\subsection{Background: UAV-Integrated VFC Architecture}
Suppose vehicles, UAVs, RSUs, and cloud servers are deployed in a VFC environment. Figure \ref{fig:vfc_framework} represents the layered architecture and communication pathways of the UAV-integrated VFC paradigm. 

\textbf{Cloud Layer:}
At the top of the paradigm lies the cloud layer, depicted with multiple cloud symbols indicating the expansive and powerful computational resources available through remote data centers. This layer represents the upper echelon of processing capability, suited for tasks requiring significant computational power and not time-sensitive.

\textbf{UAV-Integrated Cloudlet Layer:}
Below the cloud, there is the UAV-integrated cloudlet layer. This layer serves as an intermediary between the cloud and ground-level fog computing. Cloudlets are small-scale data centers that offer localized processing power, reducing latency for nearby devices. In this layer, UAVs are shown to facilitate the extension of cloudlet capabilities, implying that UAVs can carry small-scale computing infrastructure or act as communication relays to enhance the network. Both the static and mobile infrastructures are responsible for providing computation, storage, and networking services within a specific region, termed service zone in this paper.

\textbf{UAV-Integrated Fog Layer:}
On the ground level, there's the UAV-integrated fog layer, where most of the dynamic interactions occur. Here, UAVs are significantly involved in the direct processing and relaying of information. This layer showcases a rich network of connections among entities:
\begin{enumerate}
    \item Vehicles: Act as both data producers and consumers, equipped with sensors and communication capabilities.
    \item UAVs: Serve as mobile fog nodes that provide computation, storage, and networking services to nearby vehicles.
    \item RSUs: Fixed infrastructures that support communication and computation capabilities within their regions.
    \item Edge Servers: Localized servers providing processing and storage capabilities, wired to the RSUs.
\end{enumerate}
\subsection{UAV-Integrated Vehicular Fog Computing Missions}\label{sec:uvfc_research}
The UAV-integrated VFC paradigm enables a host of missions, including the task offloading, RSU/ABS deployment, UAV trajectory planning, security and privacy, and resource allocation. These missions call for diverse functionalities and operations in the simulation platform.

The joint task assignment and computation allocation to fog nodes is studied as a multi-objective minimization problem (concerning latency, energy, pricing cost, etc.), and solved via centralized or distributed methods including heuristic methods \cite{folo_zhu}, contract theory\cite{contract_matching_zhou}, matching theory\cite{contract_matching_zhou}, game theory \cite{ocvc}, reinforcement learning (RL) approaches \cite{vfc_priority_Jinming_Shi}. UAVs are also considered flexible auxiliary nodes for computation offloading in post-disaster rescue \cite{disaster_Y_Wang}. In \cite{joint_Y_Liu}, Liu \emph{et al.} studied the UAV-assisted mobile edge computing with joint communication and computation resource allocation for vehicles. 
These studies require simulation of the \emph{computation, communication, and energy models} for performance validation.

Considering the varying computational capabilities, dynamic channel state information, and reliability of moving vehicles, the task offloading missions in UAV-integrated VFC is not merely an optimization problem but intertwines with multiple dimensions of vehicular networks. Reference \cite{traffic_load_vec} focused on traffic loads in heterogeneous VFC scenarios and executed computation offloading regarding the predicted network conditions. Besides prediction-based proactive schemes, reactive methods such as redundant resource allocation and service migration\cite{redundant_resource} can also be optimized to alleviate the uncertainty in vehicular networks. The dynamics of vehicular networks and the uncertainty of computation offloading are based on models including \emph{communication channel attributes, computation queues, road topologies, and mobility}. Therefore, the simulation of both fog node network and traffic flow is a prerequisite for the task offloading mission.

As for RSU/ABS deployment issues, drones are leveraged to augment network coverage in underserved areas\cite{coverage_uav} and guarantee real-time safety of vehicles on highway\cite{highway_uav}. This mission is not without the scope of computation and communication dynamics, as the RSU/ABS deployment is closely related to the traffic conditions and the network topology. In the realm of security and attacks, research is intensifying on addressing unique cybersecurity challenges, including data privacy and secure communication\cite{se_vfc,trust_comp}. These mechanisms can be implemented by applying \emph{security operations} such as authentication, encryption, and blockchain. 
In \cite{blockchain_evastrop}, Gupta \emph{et al.} presented a blockchain-based secure scheme to prevent controller hijacking and man-in-the-middle attacks. Furthermore, the allocation of computational and communication resources among large-scale regions\cite{large_scale_vfc} and the exploration of economic models for resource sharing and trading through incentive CPU trading are gaining traction\cite{madrl}, thereby fostering a collaborative and efficient vehicular network ecosystem on the basis of incentive mechanisms.

Overall, these research efforts are jointly devoted to a secure, sustainable, collaborative, and efficient computation framework in the UAV-integrated VFC, and requires supportive functionalities to validate the propositions.

\begin{table*}[t]
    \begin{threeparttable}
    \renewcommand{\arraystretch}{1.1}
    \centering
    \caption{Comparison of Our Work with Existing Simulators}
    \label{tab:simulator_comparison}
    \begin{tabular}{||c|c|c|c|c|c|c|c|c||}
        \hline
    Software & Comm. & Comp. & En. & Sec. & Mob. & Topo. & Dependencies & Language \\ \hline
    IFogSim\cite{ifogsim} & Not Ch. att. &Yes & Yes  & No  & No & No & CloudSim\cite{cloudsim}  & Java \\
    IFogSim2\cite{ifogsim2} & Not Ch. att. &Yes & Yes  & No  & Fog node & No & CloudSim  & Java \\
    EdgeCloudSim\cite{edgecloudsim}  &Yes & Yes& No  & No & Client only & No &CloudSim & Java, Matlab \\
    FogNetSim++\cite{fognetsim++} & Yes & Yes &No & No  &  Veh. only & No & OMNeT++\cite{omenet++} & C++  \\
    Veins\cite{veins}  & Yes  & No & No  & No & Veh. only & Yes & SUMO, OMNeT++ & C++\\
    VFogSim\cite{vfogsim} & V2N only & Yes & Yes & No  & Veh. only & Yes & WinProp\cite{winprop}, SUMO & C++\\
    MARSIM\cite{marsim} & No & No & No & No & UAVs only & No & ROS & C++, C\\
    Skywalker\cite{skywalker} & Yes & Yes & Yes  & No & Veh. \& UAV & No & AnyLogic & -\\
    \textbf{AirFogSim (Ours)} &  \textbf{Yes}  & \textbf{Yes}  &  \textbf{Yes}  &  \textbf{Yes}  &  \textbf{Veh. \& UAV}  &  \textbf{Yes}  &  \textbf{SUMO}\cite{sumo_2012} & \textbf{Python}\\ \hline
    \end{tabular}
    \begin{tablenotes}
      \small 
      \item (1) ``Not Ch. att.'' means that the software does not support channel attributes (such as fading) in modeling.
      \item (2) ``Comm.'', ``Comp.'', ``En.'', ``Sec.'', ``Mob.'', ``Veh.'', and ``Topo.'' stand for communication, computation, energy, security, mobility, vehicle, and topology, respectively.
    \end{tablenotes}

    \end{threeparttable}
    \vspace{-0.4 cm}
\end{table*}
\subsection{Existing Simulation Platforms}
By delving into the current research in Section \ref{sec:uvfc_research}, the required operations for a simulation platform to manage can be summarized as follows: \emph{Communication, Computation, Energy, Security, Mobility, Traffic, and Scalability}. The scalability of the simulator is both in terms of the size of the simulation and the development of new modules. We survey the representative simulators relevant to these requirements and compare them with our proposed AirFogSim in Table \ref{tab:simulator_comparison}. 

General fog and edge computing simulators like IFogSim \cite{ifogsim}, IFogSim2 \cite{ifogsim2}, and EdgeCloudSim\cite{edgecloudsim} concentrate on computation and energy dynamics, yet they fall short in simulating critical aspects such as mobility and road topologies. Vehicular network-focused simulators, namely FogNetSim++\cite{fognetsim++} and Veins\cite{veins}, address more specialized requirements, while the former integrates computing features for complex network simulations, the latter excels in vehicular network and traffic simulations, albeit without comprehensive computation and energy components. VFogSim\cite{vfogsim} represents a category of simulators specifically tailored for VFC, with robust communication and energy modeling capabilities. However, its lack of security features is a significant gap, given the increasing cybersecurity concerns in VFC environments. In the domain of UAVs, MARSIM\cite{marsim} and Skywalker\cite{skywalker} emerge as specialized tools. MARSIM's focus on LiDAR-based UAV applications marks its niche in UAV-centric simulations, whereas Skywalker extends its utility to UAV-assisted federated computing, proving invaluable for smart city applications. However, both simulators lack the ability to simulate the urban road topology and traffic dynamics. 

While existing simulators provide valuable insights into various aspects of VFC, they exhibit notable limitations in the context of computation offloading in UAV-integrated VFC. Addressing these limitations, our proposed AirFogSim offers functionalities in Table \ref{tab:simulator_comparison}, as a modular, lightweight, and easily adaptable platform, making it a practical and efficient tool for evolving research requirements in this dynamic field. 

\begin{figure*}
    \centering
    \includegraphics[width=0.9\textwidth]{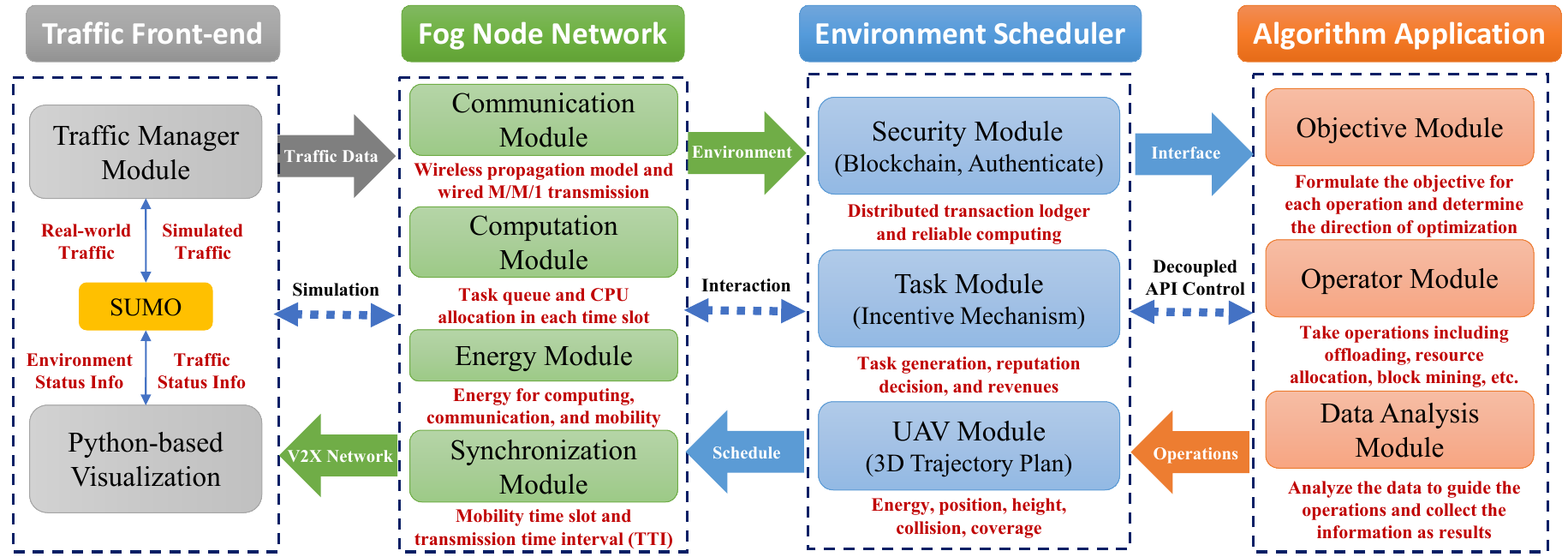}
    \caption{The system modules of the AirFogSim platform.}
    \label{fig:system_modules}
    \vspace{-0.4 cm}
\end{figure*}

\vspace{-0.2 cm}
\section{System Architecture and Module}\label{sec:system_architecture}
This section provides an overview of the proposed simulation platform, AirFogSim. 
As shown in Fig. \ref{fig:system_modules}, the proposed system architecture is stratified into four core parts: (1) Traffic Front-End, (2) Fog Network Simulation, (3) Environment Scheduler, and (4) Algorithm Application. 
\subsection{Traffic Front-End}

The ``Traffic Front-End'' serves as the foundation of the simulation environment. It utilizes SUMO (Simulation of Urban Mobility) to generate synthetic vehicular mobility patterns or utilize real-world traffic data. The integration with Python allows for the visualization of traffic scenarios, vehicular flows, UAVs, and network topologies. 
\begin{enumerate}
    \item Traffic Manager Module: Configure the traffic environment via \texttt{traci} to access SUMO. It provides the ability to generate synthetic traffic patterns or import real-world traffic data.
    \item Python-based Visualization: Interfaced with Python, this module provides an immersive visualization of traffic conditions, vehicular flows, and UAV movements. It provides both graphical via \texttt{tkinter} and tabular representations via \texttt{curses} of the simulation environment.
\end{enumerate}
\begin{figure*}[t]
    \centerline{\includegraphics[width=0.8\textwidth]{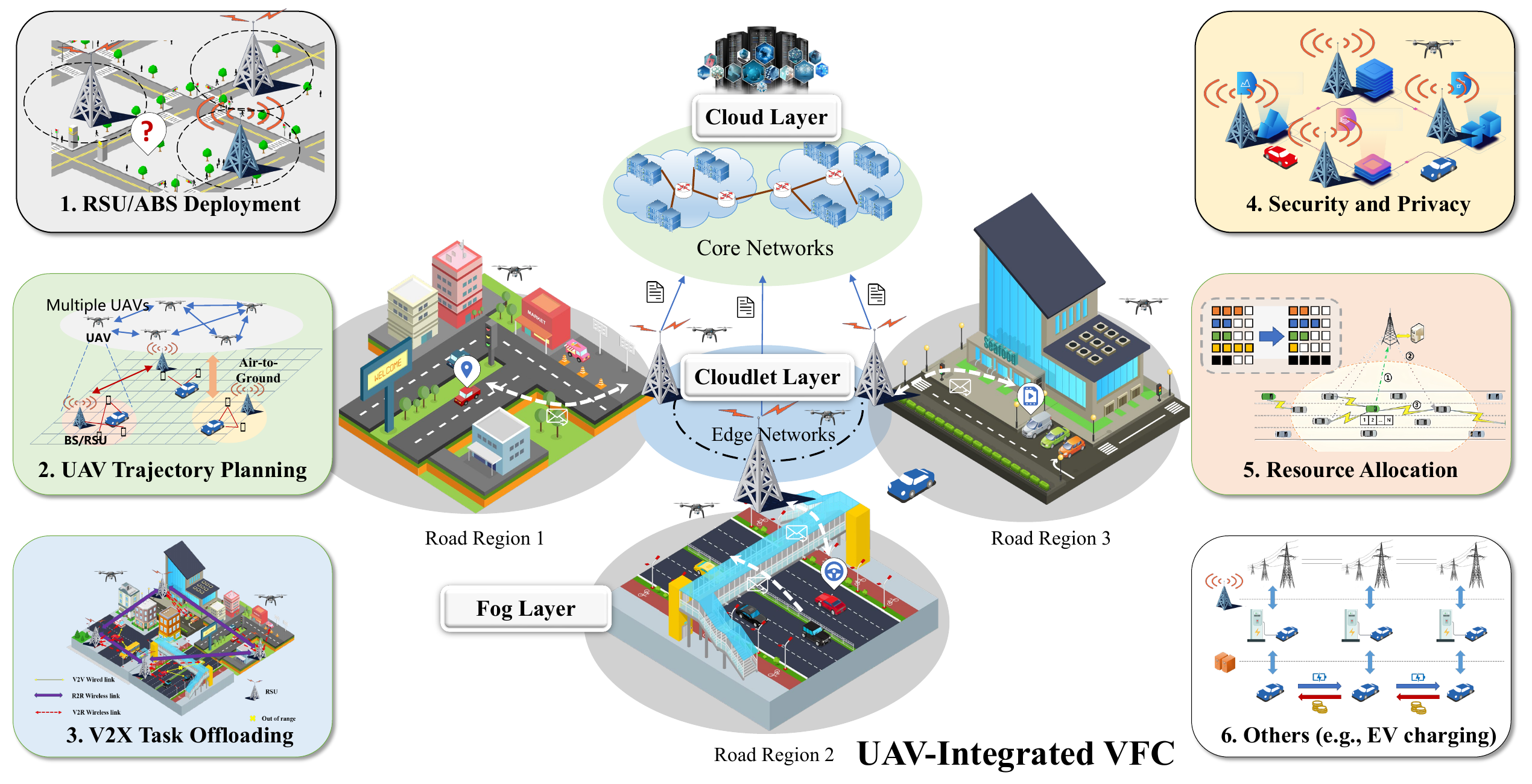}}
    \caption{Key missions supported by AirFogSim, including RSU/ABS deployment, UAV trajectory planning, V2X task offloading, security and privacy, resource allocation, etc. in the UAV-integrated VFC paradigm.}
    \label{fig:example_scenarios}
    \vspace{-0.4 cm}
\end{figure*}
\subsection{Fog Node Network}
``Fog Node Network'' epitomizes the computation, communication, and energy simulation in the environment.
\begin{enumerate}
    \item Communication Module: This module adheres to the 3GPP standards for channel propagation models. Researchers can adjust parameters online or offline to fit the dynamics of road traffic, vehicle flow, and physical obstructions that affect channel states. It offers a more lightweight and efficient solution compared with other simulators like OMNeT++\cite{omenet++} and WinProp\cite{winprop}. Wired links are also supported to simulate the communication between RSUs and cloud/edge servers via M/M/1 queues.
    \item Computation Module: This module orchestrates computational tasks across diverse fog entities. It allows for the designation of different computational sequences and CPU allocation strategies. Tasks are stored in queues and processed according to the scheduling algorithms.
    \item Energy Module: The energy module is responsible for managing the energy consumed during transmission and computation of fog entities, especially for the UAVs. This module aims to optimize energy usage across the VFC ecosystem, ensuring sustainable operation without compromising performance.
    \item Synchronization Module: This module is responsible for time synchronization. Two-time scales are supported: the simulation time scale determined by SUMO and the transmission time interval (TTI) for slot-wise computation and communication.
    It ensures that all entities are operating on the same time scale, facilitating the coordination of tasks and resources. 
\end{enumerate}
\subsection{Environment Scheduler}
The ``Environment Scheduler'' is responsible for orchestrating the simulation environment, including the operations of security, tasks, and UAVs.
\begin{enumerate}
    \item Security Module: This module leverages blockchain and authentication technologies to validate the integrity and authenticity of computation services. It incorporates verification stages to ensure the results are trustworthy. Additionally, it utilizes reputation systems to evaluate and maintain the credibility of participating entities. 
    \item Task Module: This module employs an incentive mechanism to encourage fog entities to participate in task computation and offloading processes actively. Vehicles may generate tasks and allocate resources to specific tasks according to this module. 
    \item UAV Module: This module is dedicated to optimizing the flight paths and destinations. The optimization of 3-D trajectories takes into account factors such as UAV energy consumption, computation demands, and collisions. 
\end{enumerate}

\subsection{Algorithm Application}
The uppermost ``Algorithm Application'' is the bedrock for experimentation and development. It provides a flexible framework for researchers to test and evaluate algorithms.
\begin{enumerate}
    \item Objective Function Module: This module formulates an objective function that integrates trustworthiness metrics from the ``Environment Scheduler'' to serve as the foundation for optimization. It takes a multi-criteria approach to formulate the objective of each operation.
    \item Operator Module: This module is responsible for the implementation of the optimization algorithms, including offloading, resource allocation, block mining, etc. 
    \item Data Analyses Module: This module is responsible for analyzing the data to guide the operations and collect the information as results. 
\end{enumerate}
\section{Supported Missions in AirFogSim}\label{sec:example_scenarios}
AirFogSim can support different missions thanks to the light-weight modules and multifunctional operations. In this section, we introduce the key missions in Fig. \ref{fig:example_scenarios}.
\subsection{RSU/ABS Deployment}
The RSU/ABS deployment directly influences the latency, coverage, and quality of service of the network. The objective is to optimize placement and operational parameters in line with the dynamic demands of vehicular networks and urban layouts. This mission in AirFogSim allows users to define parameters such as the number of RSUs/drones, their communication range, and processing capabilities. The simulation then proceeds to position the RSUs/drones within the virtual environment and evaluate the network performance under various traffic conditions.


\subsection{UAV Trajectory Planning}
The core objective of this mission is to develop and evaluate UAV flight strategies that minimize energy consumption and latency while maximizing the QoS and coverage. AirFogSim enables the simulation of various trajectory planning algorithms, including those based on predictive models that consider vehicular traffic patterns, urban topologies, and user demand forecasts. 

\subsection{V2X Task Offloading}
V2X task offloading is a pivotal functionality where computational tasks are transferred from vehicles to edge computing devices or cloud servers. The primary objective is to optimize the distribution of computational tasks among vehicles, UAVs, RSUs, and cloud servers to enhance operational efficiency, reduce latency, and conserve vehicular computational resources. 


\subsection{Security and Privacy}
The integration of authentication and blockchain technologies in vehicular networks introduces secure and privacy-preserving approaches to ensuring data integrity and trust among participants. The blockchain module in AirFogSim aims to simulate the process of verifying and adding transaction records to the blockchain, maintaining the ledger's reliability and security within the VFC paradigm. The authentication module, on the other hand, is responsible for verifying the identity of network entities and ensuring that only authorized users can access the system. 

\subsection{Resource Allocation}
In computation offloading, efficient allocation of communication and computation resources is crucial for the performance of a UAV-integrated VFC system. This module is dedicated to optimizing the distribution of these resources among vehicles, UAVs, and RSUs to improve overall network throughput, reduce latency, and ensure fairness. 


\subsection{Other Missions}
Researchers can further extend the AirFogSim's capabilities to explore other aspects of UAV-integrated VFC systems. For example, electric vehicle (EV) charging can be merged into VFC\cite{evfog}, which enables the simulation of EV charging stations and the allocation of charging resources to EVs at the expense of their CPU resources. This mission can be easily implemented by adding ``battery'' attributes to vehicles and developing \texttt{ChargingStation} class.

\section{Design and Implementation}\label{sec:simulator}
In this section, we introduce the platform design and implementation, including visualization, propagation modeling, computation and transmission modeling, blockchain modeling, and attack modeling.

\begin{figure}[t]
    \centering
    \includegraphics[width=0.4\textwidth]{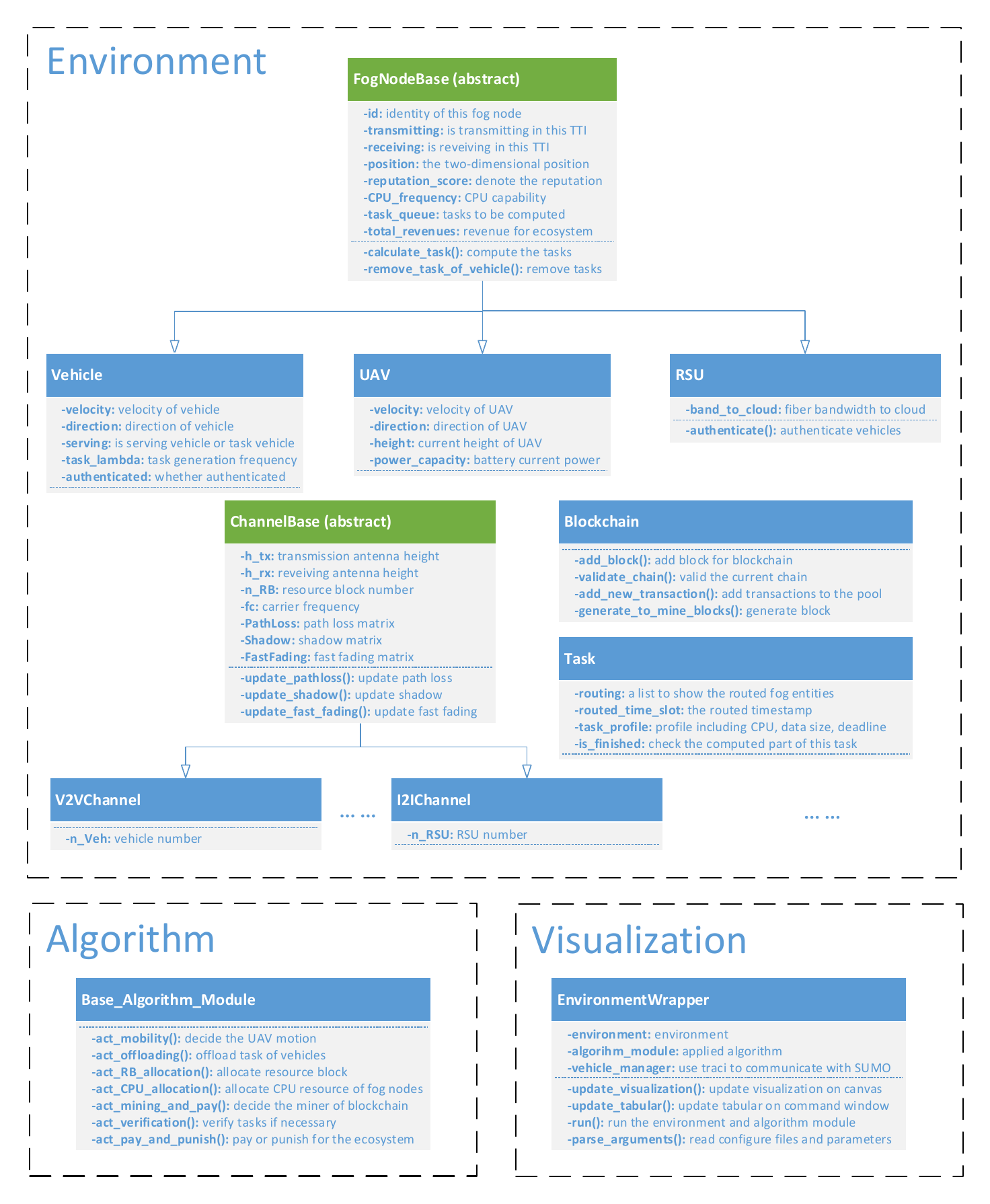}
    \caption{Several designed classes in the AirFogSim platform.}
    \label{fig:class_fig}
    \vspace{-0.4 cm}
\end{figure}
\subsection{Visualization based on SUMO Traffic}\label{subsec:sumo_traffic_gen}
Traffic flow generation serves as the foundation for simulating vehicular networks within our platform. 
We utilize the SUMO tool and the \texttt{traci} package to interactively handle large road networks and traffic flows.


A \texttt{VehicleManager} object orchestrates the generation of vehicular traffic within the simulation. It manages the introduction of individual vehicles into the traffic flow, stipulating their points of origin and intended destinations. While synthetic data is adequate for a broad range of simulation missions, the integration of real-world traffic data can provide additional verisimilitude in traffic patterns.


\subsection{Propagation Modeling}
As discussed in 3GPP Release 15 \cite{3gpp_r15} for cellular V2X enhancement, channel gain coefficients encompass the effects of frequency-independent large-scale fading (path loss, shadowing) and frequency-dependent small-scale fading (fast fading) in AirFogSim.

\subsubsection{Path Loss Model}
The path loss model for the WINNER scenarios\cite{winner2_model,3gpp_36885,3gpp_36777} is carried out as:
\begin{equation}
    PL=A\log_{10}(d)+B+C\log_{10}{\frac{f_c}{D}}
\end{equation}
where $d$ is the 3D distance between transmitter and receiver, and $f_c$ is the carrier frequency. $A,B,C,D$ are the fitting parameters where $A$ includes the path loss exponent, $B$ is the intercept, $C$ describes the path loss frequency dependence, and $D$ is the scaling factor. The fitting parameters are environment and channel-specific. For example, the WINNER+ B1 urban scenario is adopted in 3GPP TR 36.885 \cite{3gpp_36885} for V2V channels, thereby the fitting parameters are given as $A=22.7,B=41.0, C=20, D=5.0$. The path loss model can be easily changed according to practical requirements.

\subsubsection{Shadow Fading Model}
The shadow fading is assumed to initially follow the log-normal distribution with a fixed standard deviation\cite{3gpp_36885,3gpp_36777}. The shadow fading is affected by the relative movements between entities (vehicles and UAVs) based on the distances moved in the last time step. This is done using an autoregressive model where the new shadow fading is a weighted combination of the previous shadow fading and a new shadowing term. The weights depend on the relative movements between entities and the decorrelation distance. The update process can be mathematically represented as:
\begin{align}
    S_i(t+1) = &10 \cdot \log_{10} \left[\exp\left(-\frac{{\Delta d_i}}{{d_{\text{corr}}}}\right) \cdot \left(10^{\frac{{S_i(t)}}{10}}\right) + \right.\notag \\
    &\left. \sqrt{1-\exp\left(-2 \cdot \frac{{\Delta d_i}}{{d_{\text{corr}}}}\right)} \cdot 10^{\frac{{N(0, \sigma_{S_i})}}{10}}\right]
\end{align}
where $S_i(t+1)$ represents the shadow fading of entity $i$ at time step $t+1$, $\Delta d_i$ is the relative movement of entities for the $i$-th channel during the last time step, $d_{\text{corr}}$ is the decorrelation distance, $S_i(t)$ is the shadow fading at time step $t$, and $N(0, \sigma_{S_i})$ represents a normally distributed random variable with mean 0 and standard deviation $\sigma_{S_i}$. This formula allows for a dynamic update of the shadow fading as the relative positions of the entities change, thus improving the realism of the wireless signal strength simulations. 

\subsubsection{Fast Fading Model}
The fast fading is modeled as Rayleigh fading and assumed to be exponentially distributed with unit mean\cite{spectrum_vehicle_jsac_2019}. 
Hereafter, the channel power gain of the $i$-th channel can be concluded as:
\begin{equation}
    g_i^{mode}=\frac{S_i}{PL_i}h_i^{mode}
\end{equation}
where $mode$ denotes the mode of different channel frequencies, $S_i,PL_i,h_i$ are the shadow fading, path loss, and fast fading of the $i$-th channel, respectively. 

After all, we discuss the channel capacity of different link modes. Suppose that the $i$-th V2V channel is erected between vehicle $V_i$ (transmitter) and $V_j$ (receiver), the transmission rate of the $i$-th channel can be given by:
\begin{equation}\label{eq:capacity}
    C^{V2V}_{i} = x^{V2V}_{i,j} B^{V2V}_{i}\log_2{\left(1+\gamma ^{V2V}_{i}\right)}.
\end{equation}
In Eq. (\ref{eq:capacity}), $x^{V2V}_{i,j}$ is the indicator variable to show whether $V_i$ transmit to $V_j$, $B^{V2V}_{i,j}$ is the allocated bandwidth (i.e., resource blocks, RBs), and $\gamma_{i,j}^{V2V}$ is the signal-to-interference-plus-noise ratio (SINR) of the V2V communications. If the allocated RBs are shared among multiple V2V channels simultaneously, the SINR of the $i$-th V2V link is expressed as:
\begin{equation}\label{eq:sinr}
    \gamma ^{V2V}_{i}=\frac{p_i^{V2V}g^{V2V}_{i}}{N_0+\sum_{V_m, V_n\in \mathbf{Veh}, m\neq i}{x^{V2V}_{m,n}p_m^{V2V}g_m^{V2V}}}
\end{equation}
where $N_0$ is the power of complex Gaussian white noise and $g^{V2V}_i$ denotes the channel gain of the $i$-th V2V links. If the RB is occupied by only one channel, the interference disappears and the SINR $\gamma^{V2V}_i$ degenerates into SNR. 

Similarly, the transmission capacities of V2I, U2V, U2I, U2U, I2I, etc., channels can be induced by Eqs. (\ref{eq:capacity}) and (\ref{eq:sinr}).

\subsection{Computation and Transmission Modeling}\label{subsec:comp_trans_model}
This subsection elucidates the computational queuing model and the transmission scheme underpinning task offloading and execution.

\subsubsection{Task Queue Model}
For the computation model, we symbolize the task queue at any fog node \( X_j \) as \( \mathcal{I}_{X_j} \). The state of the queue at any time \( t \) can be described by the tuple \( (I_{1}^{X_j}, I_{2}^{X_j}, \ldots, I_{n}^{X_j}) \), where \( I_{i}^{X_j} \) represents the \( i \)-th task in the queue. Each task is further characterized by its own tuple \( \{X_j, up_{i}^{X_j}, req_{i}^{X_j}, \tau_{i}^{X_j}\} \), specifying the upload size, required compute cycles, and delay tolerance, respectively.

\subsubsection{CPU Resource Allocation}
In each TTI, the CPU allocation strategy is determined by the fog node's scheduling algorithm. This strategy can be modeled by adjusting the allocation of CPU resources \( \epsilon_{j,k} \) in the computation delay:

\begin{equation}
    T^{comp}_{X_j,k}=\frac{req_k}{\epsilon _{j,k}F_j}
\end{equation}
The CPU resource allocation \( \epsilon_{j,k} \) reflects the portion of the computing frequency \( F_j \) that is allocated to task \( k \) by device \( X_j \). This allocation can be dynamic and governed by various scheduling algorithms that consider factors like task urgency, resource availability, and overall system optimization goals.

\subsubsection{Transmission Model}
The spectrum is divided into many closely spaced subcarriers, which are assigned to users in a dynamic manner. The transmission delay \( T^{tran}_{i,k} \) for the \( i \)-th sub-channel, tasked with transmitting the data for vehicle \( V_k \), is inversely proportional to the sub-channel's capacity \( C_i^{mode} \):
\begin{equation}
    T^{tran}_{i,k}=\frac{up_k}{C_i^{mode}}
\end{equation}
Here, \( C_i^{mode} \) encapsulates the effects of all sub-channel bandwidth allocation, modulation scheme, and the characteristics defined by the propagation modeling.

\subsection{Blockchain Modeling}
Blockchain technology plays a pivotal role in ensuring the integrity and security of transaction data within a network. In each time slot, transactions are collected and added to a transaction pool. The blockchain modeling process can be summarized as follows:

\subsubsection{Block Generation and Mining Process} A miner is selected in accordance with the consensus algorithm and employed by the blockchain system. This miner is responsible for generating a new block, which involves collating transactions from the pool, validating them, and then broadcasting the newly created block to the network.

\subsubsection{Block Verification} Upon receipt of the new block, other nodes in the network undertake the verification process. This is a crucial step to ascertain the block's validity and to maintain the blockchain's overall consistency and reliability. Once verified, the block is appended to the blockchain, thus updating the ledger.

\subsubsection{Reward Mechanism} The miner who successfully generates a block is rewarded for their contribution to the network. This reward typically comprises two components: the transaction fees and the block reward. Transaction fees are collected from the transactions included in the block, serving as an incentive for miners to prioritize transactions with higher fees. The block reward, usually a set number of cryptocurrency units, is granted as an additional incentive for participating in the block generation process.

Currently, the supported consensus algorithm in AirFogSim is Proof-of-Stake (PoS) due to the limited onboard resource assumption of vehicles. The plan for other consensus algorithms, such as Proof-of-Work (PoW) and Proof-of-Authority (PoA), is underway.

\subsection{Attack Modeling}
Similar to previous works\cite{trust_comp,se_vfc}, three typical attacks are considered in the computation offloading of fog vehicles:
\subsubsection{Identity Spoofing Attack} In the computing ecosystem, fog nodes are rewarded by their computation. Therefore, the attacker disguises itself as a legitimate vehicle and can obtain the fees of other fog nodes. This attack can be prevented by the fog nodes' authentication mechanism.
\subsubsection{Always-On Attack} In this attack, the attacker always returns false results to the offloaded tasks to obtain the computing fees without any computation costs.
\subsubsection{On-Off Attack} Malicious fog vehicles obtain computing fees by returning correct results for a while and then returning false results so that the reputation can be maintained at a certain level. 

These three attack models can be prevented by the well-defined reputation mechanism based on the blockchain technology in the AirFogSim platform. Additional attacks and prevention methods (cipher attack, Sybil attack, etc.) will be considered in future work.


\section{Case Study: A UAV-Integrated Reliable V2X Task Offloading Framework}\label{sec:use_case}
In this section, we introduce a use case conducted by the AirFogSim platform. 
\subsection{Problem Formulation}
The communication and mobility features of vehicles and UAVs are viewed as unchanged in each time slot (i.e., TTI). We use a task set $\mathcal{I}[t]=\{I_{t,1},I_{t,2}\cdots\}$ to denote the tasks generated by the task vehicles (TVs) in each time slot $t$, where $I_{t,k}=\{V_k,up_{t,k},req_{t,k},\tau_{t,k}\}$. A set of fog nodes is denoted as $\mathcal{X}=\{X_j\}$ with computing frequencies $F^X_{j}$, including serving vehicles (SVs), UAVs, and RSUs.

The resource-constrained V2X task assignment problem coupled with UAV trajectories is then given by:
\begin{equation}\label{eq:original}
    \begin{split}
        \min\limits_{vars} \quad &\sum_{t,k} \left[T^{compE}_{t,k}+ punish\left(1-\sum_{j}\mu_{t,k,j}\right)\right]\\ 
    \text{s.t.} \quad 
    & C1:x_{t',k}[t], y_{t',k}[t], \mu_{t',k,j} \in \{0,1\}, \forall t', k, t \\ 
    & C2: \sum _{j} \mu_{t',k,j} \le 1, \forall t',k \\ 
    & C3: {x_{t',k}[t]+ y_{t',k}[t]} \le \sum _{j} \mu_{t',k,j}, \forall t',k,t \\ 
    & C4:T_{t',k}^{compE} \ge t\cdot y_{t',k}[t], \forall t',k,t:t\ge t' \\ 
    & C5.1:T_{t',k}^{compE} \ge (T-t)\cdot y_{t',k}[t], \forall t',k,t:t\ge t' \\ 
    & C5.2:T_{t',k}^{compS} \le T-T_{t',k}^{compE}, \forall t',k,t:t\ge t' \\ 
    & C6:T_{t',k}^{tranE} \ge t\cdot x_{t',k}[t], \forall t',k,t:t\ge t' \\ 
    & C7:T_{t',k}^{tranS} \le t\cdot x_{t',k}[t], \forall t',k,t:t\ge t' \\ 
    \end{split}
    \begin{split}
    & C8:T_{t',k}^{tranE} \le T_{t',k}^{compS}, \forall t',k \\ 
    & C9:T^{compE}_{t',k}-t'\le \frac{\tau_{t',k}}{dt}, \forall t',k \\  
    & C10:B[t]\cdot {\sum_{t',k}{x_{t',k}[t]}} \le B, \forall t \\ 
    & C11: F^X_j[t]\cdot {\sum_{t',k}{\mu_{t',k,j}\cdot y_{t',k}[t]}} \le {F^{X}_j} , \forall j, t \\ 
    & C12: B[t]\le B, \forall t\\
    & C13: F^X_j[t]\le F^X_j, \forall t,j\\
    & C14: \sum_{t=t':T} {x_{t',k}[t]C_{t',k,j}[t]dt} \ge  up_{t',k}\sum_{j}\mu_{t',k,j}, \forall t',k,j \\ 
    & C15: \sum_{t=t':T} {y_{t',k}[t]F^X_j[t]dt} \ge req_{t',k} \sum_{j}\mu_{t',k,j} , \forall t',k,j \\ 
    & C16: ||\delta^{uav}[t]-\delta^{uav}[t-1]||_2\leq v_{max}\cdot dt, \forall t \\ 
    \end{split}
\end{equation}
The decision variables are:
\begin{align*} &vars= \{x_{t',k}[t], y_{t',k}[t], \mu_{t',k,j}, T^{compS}_{t',k}, 
    T^{compE}_{t',k}, T^{tranS}_{t',k}, T^{tranE}_{t',k}, B[t], F^X_j[t], C_{t',k}[t], \delta^{uav}[t]\}
    \end{align*}
where $x_{t',k}[t]$, $y_{t',k}[t]$ and $\mu_{t',k,j}$ are binary decision variables indicating whether task $I_{t',k}$ is being transmitted or computed at TTI $t$ and whether it is offloaded to fog node $X_j$, respectively. $T^{compS}_{t',k}$, $T^{compE}_{t',k}$, $T^{tranS}_{t',k}$, and $T^{tranE}_{t',k}$ represent the start and end times for computation and transmission for task $I_{t',k}$. $B[t]$ is the bandwidth allocated at time slot $t$, and $F^X_j[t]$ represents the computation resources allocated by fog device $X_j$ at time slot $t$. $C_{t',k}[t]$ represents the data transmission rate at time slot $t$ for task $I_{t',k}$. $\delta^{uav}[t]$ is the position vector of UAVs at time slot $t$. $punish$ is the punishment item when a task fails to be offloaded. $dt$ is the length of time slot.

The constraints in Eq. (\ref{eq:original}) are explained as follows:
$C1$ is Binary Constraint for Decision Variables, $C2$ is Offloading Constraint, $C3$ is Transmission and Computation Constraint, $C4$ \& $C5$ are Computation Time Constraints, $C6$ \& $C7$ are Transmission Time Constraints, $C8$ is Transmission Before Computation Time Constraint, $C9$ is Delay Constraint, $C10\sim C13$ are Resource Allocation Constraints, $C14$ is Transmission Constraint, $C15$ is Computation Constraint, and $C16$ is UAV Speed Constraint. 

Due to the complexity of the original problem, we decompose it into three sub-problems: the UAV trajectory planning, V2X task assignment, and resource allocation problem. 
\subsection{Settings and Visualization}
\textbf{Methodology:} We choose a square region in Berlin, Germany, spanning 2 km $\times$ 2 km as our simulation area for its complicated road situations. The environment is equipped with two RSUs and populated by four UAVs. The RSUs are strategically positioned at the coordinates $(500,500)$ and $(1500,1500)$ within the simulation area, while the UAVs are initially randomly dispersed. The UAVs are modeled to maintain a consistent altitude of 100 m.
Detailed simulation parameters are shown in Table \ref{tab:simulation_parameters}. 

\begin{figure}[t]
    \centering
    \centering
    \subfloat[]{\includegraphics[width=3.2 in]{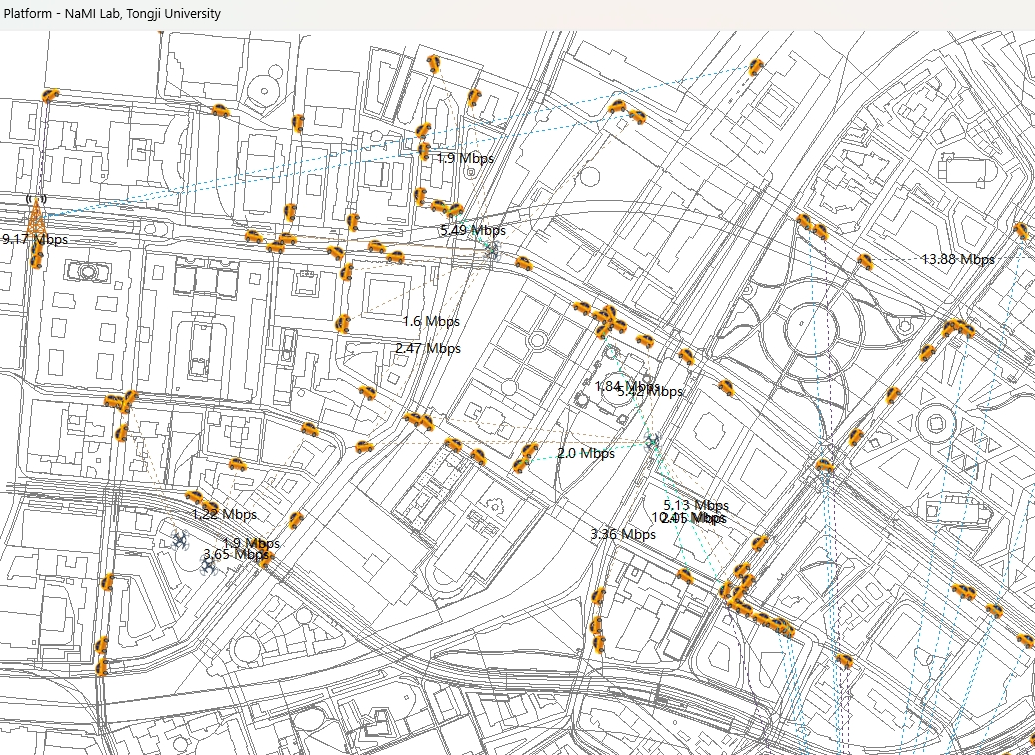}
    \label{fig:run_env}}
    \subfloat[]{\includegraphics[width=3.2 in]{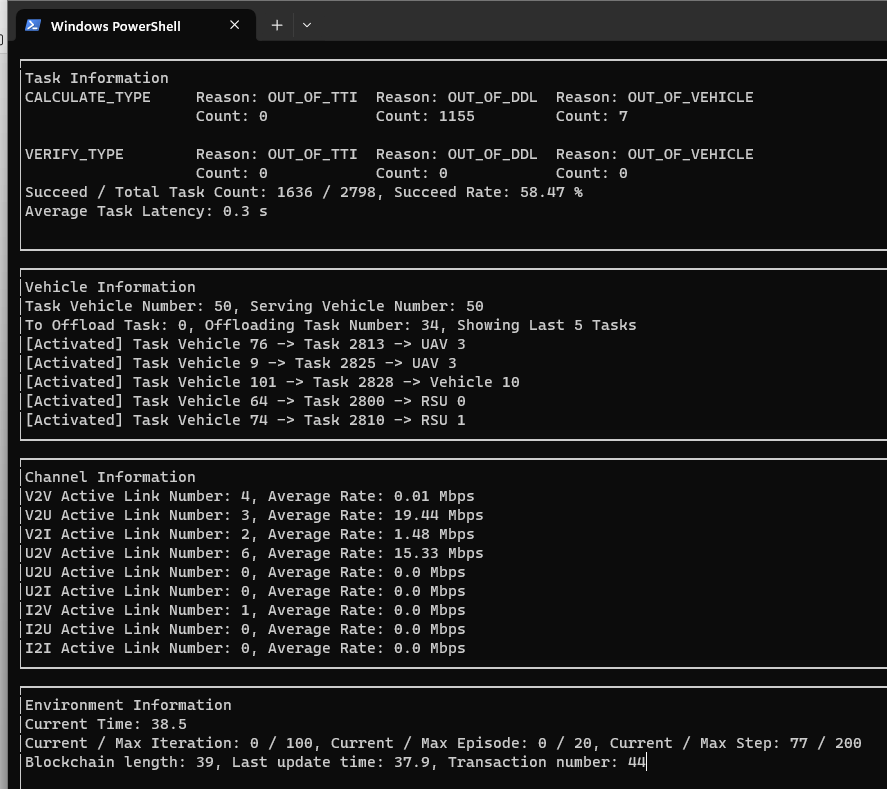}
    \label{fig:platform_visual}}
    
    \caption{Simulation environment conducted by AirFogSim. (a) graphical visualization with active wireless links and (b) tabular visualization of the simulation environment.}
    \label{fig:visualization}
    \vspace{-0.4 cm}
\end{figure}

\textbf{Simulation Results:} The visualization of our simulation environment is depicted in Fig. \ref{fig:visualization}. 

\textbf{Changed Modules:} Map and road topologies are extracted by the {OpenStreetMap} and the traffic is generated and controlled by the \texttt{VehicleManager} object. The routes of vehicles are stipulated by randomly selecting the start/end intersections.
\begin{table}[t]
    \renewcommand{\arraystretch}{1.2}
    \caption{Simulation Parameters}
    \label{tab:simulation_parameters}
    \centering
    \begin{tabular}{|c||c|}
      \hline
      UAV maximum coverage radius                  & 300 m                \\
      \hline
      RSU maximum coverage radius              & 500 m              \\
      \hline
      Maximum UAV velocity                   & 25 m/s      \\
      \hline
      V2V transmit power                   & 23 dBm      \\
      \hline
      V2U/V2R transmit power                   & 26 dBm      \\
      \hline
      U2V/U2R transmit power                   & 26 dBm      \\
      \hline
      Noise power                   & -104 dB      \\
      \hline
      Total Bandwidth                    & 20 MHz    \\
      \hline
      RB number & 20\\
      \hline
      Vehicular CPU frequency & 2.5 GHz             \\
      \hline
      UAV CPU frequency & 5 GHz             \\
      \hline
      Cloud CPU frequency & 25 GHz             \\
      \hline 
      Task deadline $\delta$                      & $[0.2,1]$ s                \\
      \hline
      Task data size                   & $[0.02,1]$ Mb \\
      \hline
      Task computational requirement $cr$     & $[0.1,0.3]$ Gigacycle        \\
      \hline
      Task vehicle lambda type                      & $[2,5,10]$ \\
      \hline
      Task vehicle lambda possibility                      & $[0.6,0.3,0.1]$ \\
      \hline
      Transmission Time Interval (TTI)     & 0.05 s        \\
      \hline
      Mobility Simulation Step     & 0.5 s        \\
      \hline
    \end{tabular}
  \end{table}

\subsection{K-Means for UAV Trajectory}
  
\begin{figure}[t]
    \centerline{\includegraphics[width=0.6\textwidth]{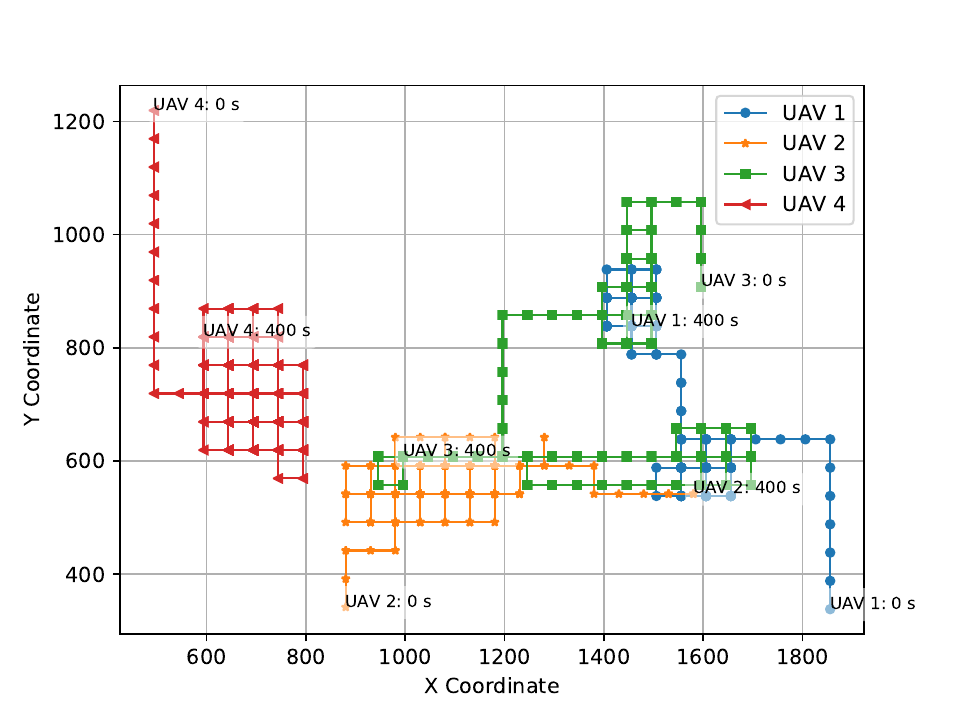}}
    \caption{Trace of 4 UAVs in the system via K-Means clustering.}
    \label{fig:kmeans_trace}
\end{figure}

\textbf{Methodology:} 
K-Means clustering algorithm is used to divide the vehicles into several clusters. The UAVs are controlled to fly to the center of the clusters to collect the data from vehicles. Then, based on the positions of UAVs and RSUs, each vehicle decides the belonging service zone by selecting the nearest zone manager. Each vehicle can only offload its tasks to the fog nodes within its service zone. 

\textbf{Simulation Results:}
In Fig. \ref{fig:kmeans_trace}, we show the trace of the four UAVs in the system. The UAVs are initially randomly dispersed in the simulation area. 

\textbf{Changed Modules:} The \texttt{act\_mobility} function in Fig. \ref{fig:class_fig} is changed to the K-Means algorithm. 

\subsection{Window-Based Hungarian for V2X Task Offloading}


    \begin{algorithm}[t]
        \SetAlgoLined
        \KwResult{Optimal $x$ and $y$}
         Initialization of $x, y$\;
         \For{$i \leftarrow 1$ \KwTo $maxIterations$}{
          Save current $x_{prev} \leftarrow x$, $y_{prev} \leftarrow y$\;
          Fix $x$, optimize $y$ by calculating the time slot allocation problem for communication resources\;
          Fix $y$, optimize $x$ by calculating the time slot allocation problem for computation resources\;
          \If{$||x - x_{prev}||_2 < tolerance$ \textbf{and} $||y - y_{prev}||_2 < tolerance$}{
           \textbf{break}\;
           }
         }
         \caption{Alternating Optimization Algorithm}\label{alg:ao}
        \end{algorithm}

\textbf{Methodology:} The Hungarian algorithm, also known as the Kuhn-Munkres algorithm, is a combinatorial optimization algorithm that solves the assignment problem in polynomial time. For our scenario, the problem can be defined as finding the best assignment of tasks to devices that minimizes the total cost within the time window $ws$ (defined as 10 TTIs in this part). The cost can be interpreted as the resource consumption or delay incurred by assigning a specific task to a particular device. When a task is offloaded to a device at time slot $t$, the task must be completed within the duration $t+ws$. 

\textbf{Simulation Results:} The comparison of latency, complexity, and successful ratio is shown in Table \ref{tab:roo_who_comparison} and Fig. \ref{fig:snum_metrics}. Detailed analyses of the results are presented in Section \ref{sec:res_alloc}.

\textbf{Changed Modules:} In the algorithm application module, the \texttt{act\_offloading} function is changed to the window-based Hungarian algorithm.

\subsection{Alternating Optimization for Resource Allocation}\label{sec:res_alloc}
\textbf{Methodology:}
Given the fixed task assignment relationship, the time slot allocation problem of joint communication and computation resources is formulated as a mixed-integer linear programming (MILP) problem, which is NP-hard. However, we find that the discrete variables in time slot allocation can be relaxed as continuous variables respectively for the transmission slot allocation problem and computing slot allocation problem. It is mainly because the optimal solution of the MILP problem equals to the one of the LP problem if the time-sharing condition holds\cite{time_sharing_WC_2006}. Therefore, it leaves room for iterative optimization of the two sub-problems. 

In this part, we propose an innovative approach that relies on the principles of Alternating Optimization (AO) to mitigate such complexities. Our AO-based two-step strategy breaks down the original problem into manageable sub-problems, each focusing on either the transmission time or the computational resource allocation. The AO-based strategy is shown in Algorithm \ref{alg:ao}. The result of Algorithm \ref{alg:ao} is the optimal $x$ and $y$ that decides the time slot allocation decisions. The AO-based strategy is guaranteed to converge to a global optimum since the two slot allocation problems are convex given the fixed task assignment matrix. 
\begin{table}[t]
    
    \renewcommand{\arraystretch}{1.2}
    \centering
    \caption{Comparison for the Gurobi, WHO, and Greedy methods}\label{tab:roo_who_comparison}
    \begin{threeparttable}
    \begin{tabular}{cccc}
    \hline
    \textbf{SV/TV} & \textbf{Latency (s)} & \textbf{Ratio (\%)}  & \textbf{Complexity (s)} \\
    \hline
    (5, 5) & \textbf{0.133, 0.133, 0.133} & \textbf{100, 100, 100} &257, 0.018, 0.007\\
    (10, 5) & \textbf{0.133, 0.133}, 0.554& \textbf{100, 100}, 76.6 &500, 0.031, 0.012\\
    (5, 10) & -, 0.524, 0.771 & -, 78.7, 65.3 & -, 0.048, 0.015\\
    (10, 10) &  -, 0.179, 0.589 & -, 97.3, 74.7 & -, 0.064, 0.025\\
    \hline
    \end{tabular}
    \end{threeparttable}
\end{table}
\begin{figure}[t]
    \centering
    \centering
    \subfloat[]{\includegraphics[width=0.4\textwidth]{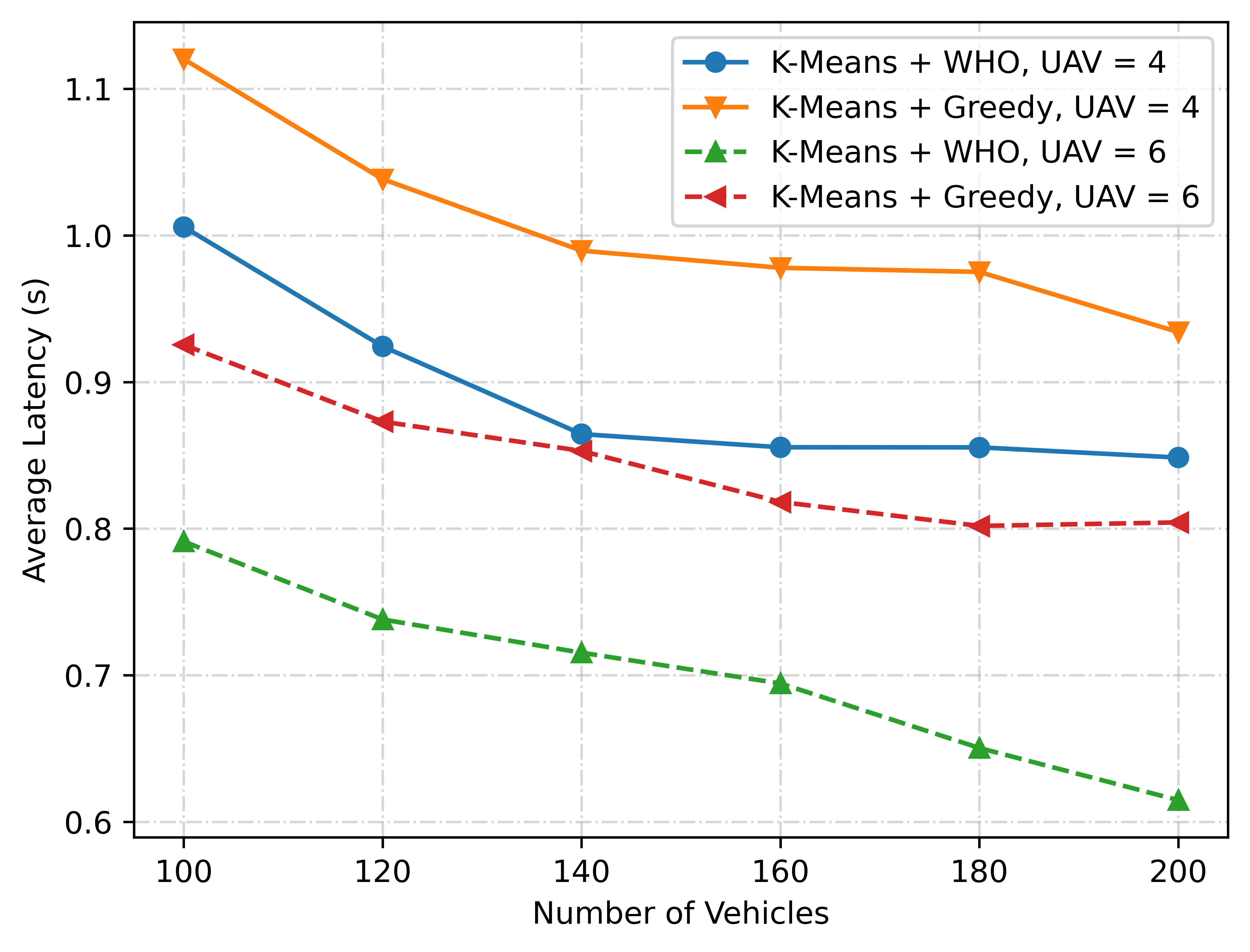}
    \label{fig:snum_qos}}
    \subfloat[]{\includegraphics[width=0.4\textwidth]{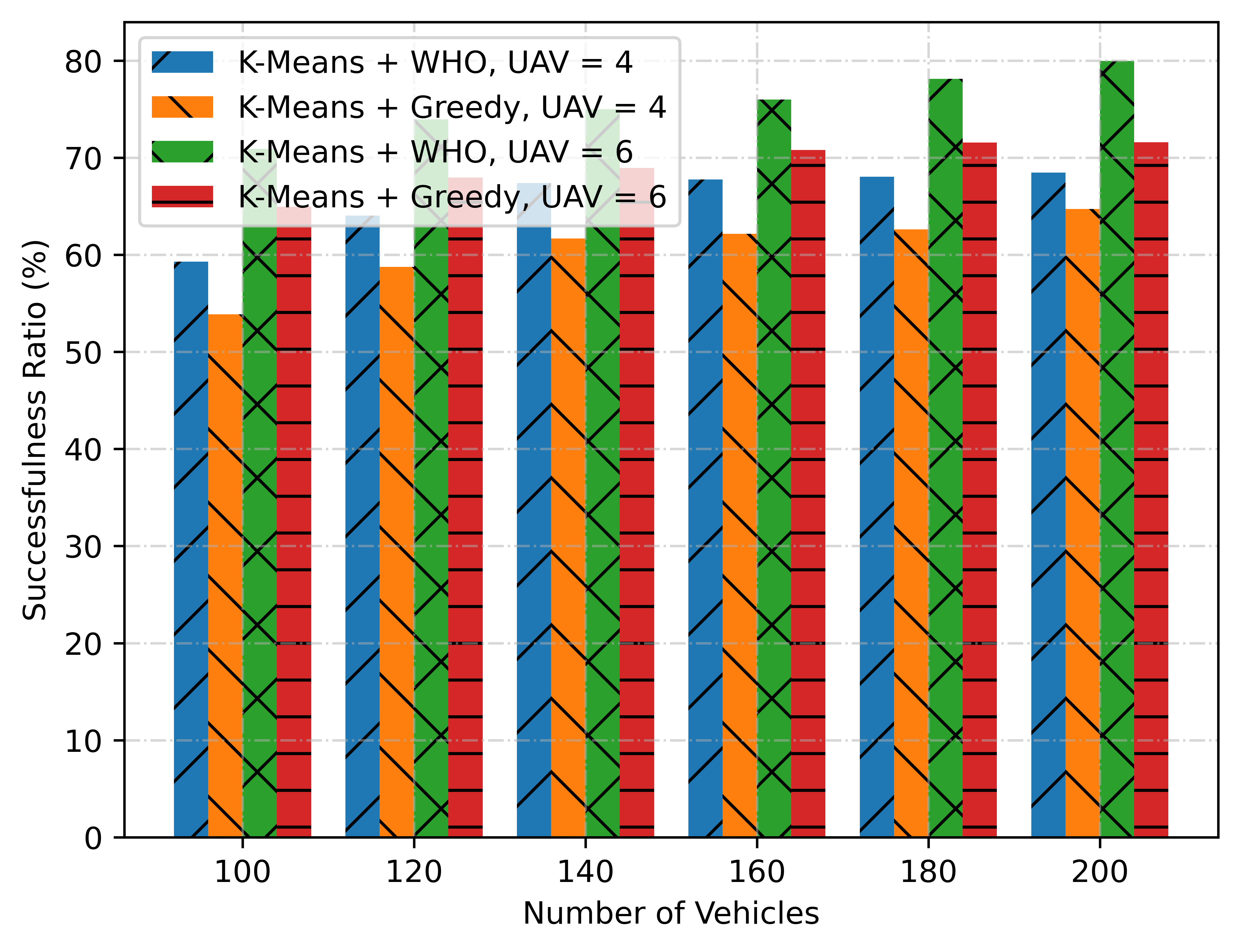}
    \label{fig:snum_ratio}}
    
    \caption{The performance comparison of different situations with fixed 50 task vehicles. (a) The average latency, and (b) successful ratio.}
    \label{fig:snum_metrics}
\end{figure}

\textbf{Simulation Results:}
As shown in Table \ref{tab:roo_who_comparison}, we elaborate on the WHO (joint window-based Hungarian and AO), the Gurobi solver (commercial solver whose solutions are deemed as ground truth), and the greedy optimization in a small-scale area. From Table \ref{tab:roo_who_comparison}, we can observe that the WHO method achieves similar performance as the Gurobi solver, while the WHO method has much lower computational complexity. When the numbers of SV and TV reach 10, the complicated task offloading problem cannot be solved within 500 seconds, which is extremely impractical in real-world scenarios. The WHO method performs a good trade-off between performance and complexity.

In Fig. \ref{fig:snum_metrics}, the number of TVs is fixed at 50, and the varying number of SVs reflect different serving situations. When the number of UAVs is 4 and the number of total vehicles is less than 140, the latency of WHO drops sharply compared with the greedy algorithm in Fig. \ref{fig:snum_qos}, which indicates a better utilization of the fog nodes' computation capabilities. When the number of vehicles becomes larger, the average latency is then constrained by the communication costs, with the successful ratio of 68\% in Fig. \ref{fig:snum_ratio}.
However, when the number of UAVs changes to 6, the aerial-to-ground communication costs are reduced, and the average latency is also reduced. The successful ratio of WHO grows to 80\% steadily with the increment of SV number in Fig. \ref{fig:snum_ratio}, while the one of the greedy algorithm is 72\%.

\textbf{Changed Modules:} In the algorithm application module, the \texttt{act\_RB\_allocation} and \texttt{act\_CPU\_allocation} functions are changed. In detail, the solution is derived by the AO-based strategy in Algorithm \ref{alg:ao}, and the resource allocation results are stored as properties in the algorithm module, returned directly when calling these two functions.

\subsection{Proof-of-Stake for Blockchain Mining}
\textbf{Methodology:}
We propose a blockchain-enabled framework that records every offloading transaction, thus ensuring an immutable and transparent task management process. 
Consensus mechanisms play a vital role in the integrity of the blockchain. Given the computational intensity and energy inefficiency of PoW, our approach adopts the PoS consensus algorithm. This method offers a more sustainable alternative, significantly reducing the energy footprint by selecting validators (namely RSUs) based on the number of tokens they hold and are willing to stake. Each recorded transaction within the blockchain encapsulates the crucial details of the offloading event: the identifier of the paying vehicle, the receiving fog nodes, the monetary amount, and the associated task profiles, including computational requirements and expected execution time frames. The block generation policy in our framework is twofold: blocks are produced at fixed time intervals of one second, and a new block is initiated once the transaction count reaches a threshold of one hundred, thereby balancing timeliness with transactional throughput.

\begin{figure}[t]
    \centerline{\includegraphics[width=0.6\textwidth]{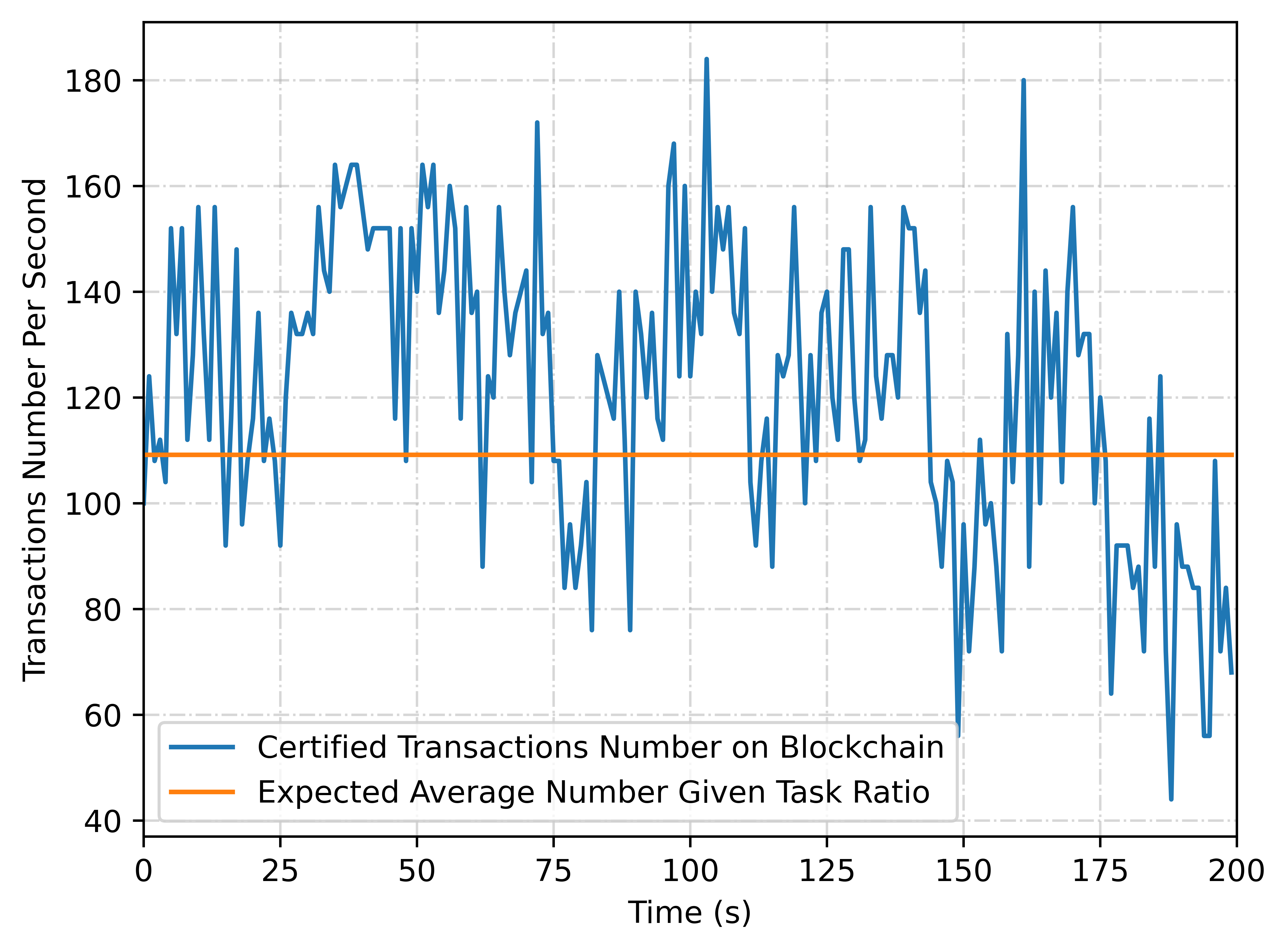}}
    \caption{Transaction number per second with fixed 50 task vehicles, 50 serving vehicles, and 4 UAVs along the simulation time.}
    \label{fig:snum_trans_num}
    \vspace{-0.4 cm}
\end{figure}
\textbf{Simulation Results:}
As shown in Fig. \ref{fig:snum_trans_num}, the certificated transaction number per second on blockchain fluctuates around 110, which is the expected value given task completion ratio as 58.9 \% in Fig. \ref{fig:snum_ratio} when the numbers of serving vehicles and task vehicles are 50. The transaction number per second is stable and reliable, which is the basis of the blockchain-enabled task offloading system. 

\textbf{Changed Modules:} In the algorithm application module, the \texttt{act\_mining\_and\_pay} function is changed to the PoS consensus algorithm. Then, the \texttt{act\_pay\_and\_punish} is changed to generate the transactions in the blockchain according to the computation results in each time slot.

\section{Conclusion and Future Work}\label{sec:conclusion}
In this paper, we presented AirFogSim, a simulation platform that contributes to addressing the challenges of computation offloading in UAV-integrated VFC. Compared with current simulators, the proposed AirFogSim offers a more comprehensive and realistic simulation environment, focusing on the unique characteristics of UAVs and VFC in multiple layers, and providing several key missions in this field.
We also demonstrated the capabilities of AirFogSim through a case study of computation offloading in VFC. The results show that AirFogSim can effectively simulate the complex interactions between UAVs and vehicles.

Future work includes enriching AirFogSim with more diverse missions and robust security models and applying the platform to a broader range of applications in ITS. Our aim is to continuously refine AirFogSim, making it an increasingly effective tool for the research community and contributing to the evolution of intelligent transportation systems.


\bibliographystyle{unsrt}  


\begin{thebibliography}{1}
    \bibitem{wireless_era_X_Cheng}X. Cheng, R. Zhang, and L. Yang, ``Wireless Toward the Era of Intelligent Vehicles,'' \emph{IEEE Internet of Things Journal}, vol. 6, no. 1, pp. 188-202, Feb. 2019.
    \bibitem{blog_cavdata} Tuxera, "Autonomous cars – the data storage challenge," Tuxera Blog, [Online]. Available: https://www.tuxera.com/blog/autonomous-cars-300-tb-of-data-per-year/. [Accessed: Nov. 28, 2023].
    \bibitem{AIGC_vehicular_metaverse_Xu}Xu, Minrui, \emph{et al.} "Generative AI-Empowered Simulation for Autonomous Driving in Vehicular Mixed Reality Metaverses." arXiv preprint arXiv:2302.08418 (2023).
    \bibitem{foggy_Muham}M. A. U. Rehman, M. Salah ud din, S. Mastorakis, and B. -S. Kim, ``FoggyEdge: An Information-Centric Computation Offloading and Management Framework for Edge-Based Vehicular Fog Computing,'' \emph{IEEE Intelligent Transportation Systems Magazine}, vol. 15, no. 5, pp. 78-90, Sept.-Oct. 2023.

    \bibitem{folo_zhu} C. Zhu \emph{et al.}, ``Folo: Latency and Quality Optimized Task Allocation in Vehicular Fog Computing,'' \emph{IEEE Internet of Things Journal}, vol. 6, no. 3, pp. 4150-4161, Jun. 2019.

    \bibitem{contract_matching_zhou}Z. Zhou \emph{et al.}, ``Computation Resource Allocation and Task Assignment Optimization in Vehicular Fog Computing: A Contract-Matching Approach,'' \emph{IEEE Transactions on Vehicular Technology}, vol. 68, no. 4, pp. 3113-3125, Apr. 2019.
  
    \bibitem{ocvc}Z. Wei, B. Li, R. Zhang, X. Cheng, and L. Yang, ``OCVC: An Overlapping-Enabled Cooperative Vehicular Fog Computing Protocol,'' \emph{IEEE Transactions on Mobile Computing}, vol. 22, no. 12, pp. 7406-7419, Dec. 2023.
    
    \bibitem{vfc_priority_Jinming_Shi}J. Shi, J. Du, J. Wang, J. Wang, and J. Yuan, ``Priority-Aware Task Offloading in Vehicular Fog Computing Based on Deep Reinforcement Learning,'' \emph{IEEE Transactions on Vehicular Technology}, vol. 69, no. 12, pp. 16067-16081, Dec. 2020.

    \bibitem{se_vfc}X. Liu, W. Chen, Y. Xia, and C. Yang, ``SE-VFC: Secure and Efficient Outsourcing Computing in Vehicular Fog Computing,'' \emph{IEEE Transactions on Network and Service Management}, vol. 18, no. 3, pp. 3389-3399, Sept. 2021.

    \bibitem{large_scale_vfc}Y. Hou, Z. Wei, R. Zhang, X. Cheng, and L. Yang, ``Hierarchical Task Offloading for Vehicular Fog Computing Based on Multi-Agent Deep Reinforcement Learning,'' \emph{IEEE Transactions on Wireless Communications}.

    \bibitem{madrl}Z. Wei, B. Li, R. Zhang, X. Cheng, and L. Yang, ``Many-to-Many Task Offloading in Vehicular Fog Computing: A Multi-Agent Deep Reinforcement Learning Approach,'' \emph{IEEE Transactions on Mobile Computing}.

    \bibitem{trust_comp}S. Xu, C. Guo, R. Q. Hu, and Y. Qian, ``Blockchain-Inspired Secure Computation Offloading in a Vehicular Cloud Network,'' \emph{IEEE Internet of Things Journal}, vol. 9, no. 16, pp. 14723-14740, 15 Aug.15, 2022.

    \bibitem{traffic_load_vec}A. Bozorgchenani, S. Maghsudi, D. Tarchi, and E. Hossain, ``Computation Offloading in Heterogeneous Vehicular Edge Networks: On-Line and Off-Policy Bandit Solutions,'' \emph{IEEE Transactions on Mobile Computing}, vol. 21, no. 12, pp. 4233-4248, 1 Dec. 2022.

    \bibitem{redundant_resource}A. S. Shafigh, B. Lorenzo, S. Glisic, and Y. Fang, ``Low-Latency Robust Computing Vehicular Networks,'' \emph{IEEE Transactions on Vehicular Technology}, vol. 72, no. 2, pp. 2130-2144, Feb. 2023.

    \bibitem{coverage_uav}M. Samir, D. Ebrahimi, C. Assi, S. Sharafeddine, and A. Ghrayeb, ``Leveraging UAVs for Coverage in Cell-Free Vehicular Networks: A Deep Reinforcement Learning Approach,'' \emph{IEEE Transactions on Mobile Computing}, vol. 20, no. 9, pp. 2835-2847, 1 Sept. 2021.

    \bibitem{highway_uav} J. Li, X. Cao, D. Guo, J. Xie, and H. Chen, ``Task Scheduling With UAV-Assisted Vehicular Cloud for Road Detection in Highway Scenario,'' \emph{IEEE Internet of Things Journal}, vol. 7, no. 8, pp. 7702-7713, Aug. 2020.
    
    \bibitem{disaster_Y_Wang}Y. Wang \emph{et al.}, ``Task Offloading for Post-Disaster Rescue in Unmanned Aerial Vehicles Networks,'' \emph{IEEE/ACM Transactions on Networking}, vol. 30, no. 4, pp. 1525-1539, Aug. 2022.

    \bibitem{joint_Y_Liu}Y. Liu \emph{et al.}, ``Joint Communication and Computation Resource Scheduling of a UAV-Assisted Mobile Edge Computing System for Platooning Vehicles,'' \emph{IEEE Transactions on Intelligent Transportation Systems}, vol. 23, no. 7, pp. 8435-8450, Jul. 2022. 
    
    \bibitem{blockchain_evastrop}R. Gupta, M. M. Patel, S. Tanwar, N. Kumar, and S. Zeadally, ``Blockchain-Based Data Dissemination Scheme for 5G-Enabled Softwarized UAV Networks,'' \emph{IEEE Transactions on Green Communications and Networking}, vol. 5, no. 4, pp. 1712-1721, Dec. 2021.




    \bibitem{ifogsim}Gupta, Harshit, \emph{et al.} ``iFogSim: A toolkit for modeling and simulation of resource management techniques in the Internet of Things, Edge and Fog Computing Environments.'' \emph{Software: Practice and Experience} 47.9 (2017): 1275-1296.

    \bibitem{ifogsim2}Mahmud, Redowan, \emph{et al.} ``iFogSim2: An Extended iFogSim Simulator for Mobility, Clustering, and Microservice Management in Edge and Fog Computing Environments.'' \emph{Journal of Systems and Software} 190 (2022): 111351.

    \bibitem{edgecloudsim}Sonmez, Cagatay, Atay Ozgovde, and Cem Ersoy. ``Edgecloudsim: An Environment for Performance Evaluation of Edge Computing Systems,'' \emph{Transactions on Emerging Telecommunications Technologies} 29.11 (2018): e3493.

    \bibitem{fognetsim++}T. Qayyum, A. W. Malik, M. A. Khan Khattak, O. Khalid, and S. U. Khan, ``FogNetSim++: A Toolkit for Modeling and Simulation of Distributed Fog Environment,'' \emph{IEEE Access}, vol. 6, pp. 63570-63583, 2018.

    \bibitem{veins} C. Sommer, R. German, and F. Dressler, ``Bidirectionally Coupled Network and Road Traffic Simulation for Improved IVC Analysis,'' \emph{IEEE Transactions on Mobile Computing}, vol. 10, no. 1, pp. 3-15, Jan. 2011.
    \bibitem{vfogsim}Ö. U. Akgül, W. Mao, B. Cho, and Y. Xiao, ``VFogSim: A Data-Driven Platform for Simulating Vehicular Fog Computing Environment,'' \emph{IEEE Systems Journal}, vol. 17, no. 3, pp. 5002-5013, Sept. 2023.

    \bibitem{marsim}F. Kong \emph{et al.}, ``MARSIM: A Light-Weight Point-Realistic Simulator for LiDAR-Based UAVs,'' \emph{IEEE Robotics and Automation Letters}, vol. 8, no. 5, pp. 2954-2961, May. 2023.

    \bibitem{skywalker}K. Hayawi, Z. Anwar, A. W. Malik, and Z. Trabelsi, ``Airborne Computing: A Toolkit for UAV-Assisted Federated Computing for Sustainable Smart Cities,'' \emph{IEEE Internet of Things Journal}, vol. 10, no. 21, pp. 18941-18950, 1 Nov.1, 2023.
    
    \bibitem{3gpp_r15} 3GPP, ``Technical Specification Group Radio Access Network; Study Enhancement 3GPP Support for 5G V2X Services; (Release 15),'' Technical Specification (TS) TR 22.886, {3rd Generation Partnership Project (3GPP)}, Mar. 2017.
    \newblock Version 15.3.0.

    \bibitem{winner2_model} P.~Kysti, J.~Meinil, L.~Hentil, X.~Zhao, and T.~Rautiainen, ``IST-4-027756 Winner II D1.1.2 v1.2 Winner II Channel Models,'' 2008.

    \bibitem{3gpp_36885} 3GPP, ``Technical Specification Group Radio Access Network; Study LTE-Based V2X Services; (Release 14),'' Technical Specification (TS) 36.885, {3rd Generation Partnership Project (3GPP)}, Jun. 2016.
    \newblock Version 14.0.0.

    \bibitem{3gpp_36777} {3GPP}, ``Technical Specification Group Radio Access Network; Study on Enhanced LTE Support for Aerial Vehicles; (Release 15),'' Technical Specification (TS) 36.777, {3rd Generation Partnership Project (3GPP)}, Dec. 2017.
    \newblock Version 15.0.0.

    \bibitem{cloudsim}Calheiros, Rodrigo N., \emph{et al.} ``CloudSim: A Toolkit for Modeling and Simulation of Cloud Computing Environments and Evaluation of Resource Provisioning Algorithms.'' \emph{Software: Practice and experience} 41.1 (2011): 23-50.

    \bibitem{omenet++}A. Varga and R. Hornig, ``An Overview of the OMNeT++ Simulation Environment,'' in \emph{Proc. 1st Int. Conf. Simul. Tools Techn. Commun., Netw. Syst. Workshops}, 2008, pp. 1–10.
    
    \bibitem{sumo_2012}D.~Krajzewicz, J.~Erdmann, M.~Behrisch, and L.~Bieker, ``Recent Development and Applications of SUMO - Simulation of Urban MObility,'' {\em International Journal on Advances in Systems \& Measurements}, 2012.

    \bibitem{winprop}R. Hoppe, G. Wölfle, and U. Jakobus, ``Wave Propagation and Radio Network Planning Software WinProp Added to the Electromagnetic Solver
    Package FEKO,'' in \emph{Proc. Int. Appl. Comput. Electromagn. Soc. Symp.}, 2017, pp. 1–2.x

    \bibitem{evfog}Z. Wei, B. Li, R. Zhang, and X. Cheng, ``Contract-Based Charging Protocol for Electric Vehicles With Vehicular Fog Computing: An Integrated Charging and Computing Perspective,'' \emph{IEEE Internet of Things Journal}, vol. 10, no. 9, pp. 7667-7680, 1 May., 2023.
    
    \bibitem{spectrum_vehicle_jsac_2019} L.~Liang, H.~Ye, and G.~Y. Li, ``Spectrum Sharing in Vehicular Networks Based on Multi-Agent Reinforcement Learning,'' {\em IEEE Journal on Selected Areas in Communications}, vol.~37, no.~10, pp.~2282--2292, 2019.

    
    \bibitem{time_sharing_WC_2006} W.~Yu and R.~Lui, ``Dual Methods for Nonconvex Spectrum Optimization of Multicarrier Systems,'' {\em IEEE Transactions on Communications}, vol.~54, no.~7, pp.~1310--1322, 2006.
\end{thebibliography}

\end{document}